# Critical temperature of superconducting bilayers: theory and experiment

G. Brammertz[1], A.A. Golubov[2], P. Verhoeve[1], R. den Hartog[1], T. Peacock[1], H. Rogalla[2]

[1] Science Payloads Technology Division, Research and Science Support Department, European Space Agency, ESTEC, Keplerlaan 1, 2200 AG Noordwijk, The Netherlands.
[2] Department of Applied Physics, University of Twente, P.O. Box 217, 7500 AE Enschede, The Netherlands.

A generalized model for the critical temperature $T_c$ of superconducting bilayers is presented, which is valid with no restrictions to film thicknesses, $T_c$ of the layers and interface resistivity. The model is verified experimentally on a series of Nb-Al and Ta-Al bilayers with Nb, Ta layer thicknesses of 100 nm and Al layer thicknesses ranging from 5 nm to 200 nm. Excellent agreement between theory and experiment was found for the energy gap and the $T_c$ of bilayers. The results are important for designing practical superconducting devices.

There is presently a growing interest in the development of superconducting bilayers $S_1$-$S_2$ for a number of practical applications, like tunnel junctions for X-ray detection[1-4], transition edge sensors[1-2,5] and Josephson junctions[6]. When designing such devices, it is often necessary to adjust and predict the bilayer transition temperature $T_c$. Although the proximity effect theory has been extensively developed during the last decade, there was certainly a lack of practically oriented studies of $T_c$ in bilayers. In early work on the proximity effect[7-10] the approximate methods for the determination of $T_c$ were developed. However, the boundary conditions used do not follow from the microscopic theory of superconductivity, see also review[11]. More recently, the microscopic theory in the dirty limit based on the Usadel equations[12] was used to calculate $T_c$ on bi- and multi-layered systems, but only limiting cases were studied: high transparency of $S_1$-$S_2$ interface[13,14], finite transparency of $S_1$-$S_2$ interface but thin or thick $S_2$ layers and zero $T_{cS_2}$[15,16] or the limit of very thin $S_1$ and $S_2$ layers[17,18].

In this paper the generalized model for $T_c$ of superconducting bilayers is presented for the first time without restriction to the $S_1$ and $S_2$ layer thicknesses, $T_{cS_1,S_2}$ values, material parameters and resistivity of the $S_1$-$S_2$ interface. The model is verified experimentally on a series of Nb-Al and Ta-Al bilayers.

We consider a bilayer structure consisting of $S_1$ and $S_2$ layers of thickness $d_{S_1}$ and $d_{S_2}$ respectively. Finite transparency of the $S_1$-$S_2$ boundary can result either from a difference in Fermi velocities of the materials or from the existence of a potential barrier at the interface. In general the $S_2$ material is also superconducting with a transition temperature $T_{cS_2} < T_{cS_1}$. Our assumption is that the materials are either in the dirty limit or in the clean limit with, in that case only, the additional condition of diffusive scattering at the film interfaces. Under these assumptions and in the vicinity of the transition temperature $T_c$ the proximity effect is described by the linearized set of Usadel equations[12] in material i=1,2:

$$\Phi_{S_i} = \Delta_{S_i} + \frac{D_{S_i}}{2\omega_n G_{S_i}}\left[G_{S_i}^2 \Phi'_{S_i}\right]', \qquad (1)$$

$$\Delta_{S_i} \ln\frac{T_c}{T_{cS_i}} + 2\pi T_c \sum_{\omega_n>0} \frac{(\Delta_{S_i} - \Phi_{S_i} G_{S_i})}{\omega_n} = 0, \qquad (2)$$

$\Phi_{S_i}$ and $G_{S_i} = \omega_n/(\omega_n^2 + \Phi_{S_i}^2)^{1/2}$ are the normal and anomalous Green's functions, $\Delta_{S_i}$ is the order parameter, and $D_{S_i}$ is the diffusion coefficient in the $S_1$ and $S_2$ layers respectively. $\omega_n = \pi T(2n+1)$ is the Matsubara frequency. Eqs. (1) and (2) are supplemented with the boundary conditions at the free surfaces, $\Phi'_{S_1}=0$ at $x=d_{S_1}$, $\Phi'_{S_2}=0$ at $x=-d_{S_2}$, and at the NS interface[19] (x=0):

$$D_{S_1}^{1/2} G_{S_1}^2 \Phi'_{S_1} = \gamma D_{S_2}^{1/2} G_{S_2}^2 \Phi'_{S_2}, \text{ and} \qquad (3)$$

$$\gamma_{BN} \xi_{S_2}^* G_{S_2}^2 \Phi'_{S_2} = G_{S_1}(\Phi_{S_1} - \Phi_{S_2}), \qquad (4)$$

where $\xi_{S_2}^* = \xi_{S_2}\sqrt{T_{cS_2}/T_{cS_1}}$ is the normalized coherence length in the $S_2$ layer, $\xi_{S_i} = (D_{S_i}/2\pi T_{cS_i})^{1/2}$. Here, the dimensionless parameters $\gamma$ and $\gamma_{BN}$ describe the nature of the interface between the two materials. They are defined by:

$$\gamma = \frac{\rho_{S_1}\xi_{S_1}}{\rho_{S_2}\xi_{S_2}^*}, \quad \gamma_{BN} = \frac{R_B}{\rho_{S_2}\xi_{S_2}^*}, \qquad (5)$$

where $\rho_{S_1}$ and $\rho_{S_2}$ are the normal state resistivities and $R_B$ is the product of the resistance of the $S_1$-$S_2$ boundary and its area. Replacing the coherence length in the films by the dirty limit expression $\xi = \sqrt{\xi_0 l/3}$, where $\xi_0$ is the coherence length in the bulk material and $l$ is the mean free path in the film, yields:

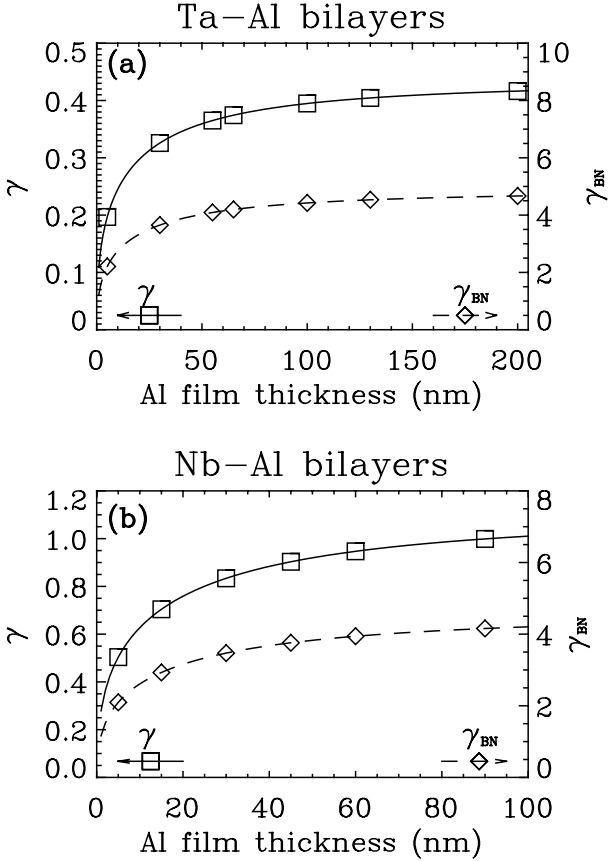

FIG. 1. Interface parameters $\gamma$ (solid line, left scale) and $\gamma_{BN}$ (dashed line, right scale) as a function of Al film thickness for **(a)** Ta-Al and **(b)** Nb-Al bilayers. Squares ($\gamma$, left scale) and diamonds ($\gamma_{BN}$, right scale) indicate the points for which calculations of the energy gap and $T_c$ were made.

$$\gamma = C_\gamma \sqrt{\frac{l_{S_2}}{l_{S_1}}}, \text{ with } C_\gamma = \frac{\rho_{S_1} l_{S_1}}{\rho_{S_2} l_{S_2}} \sqrt{\frac{\xi_{0S_1} T_{cS_1}}{\xi_{0S_2} T_{cS_2}}}, \quad (6)$$

$$\gamma_{BN} = C_{\gamma_{BN}} \sqrt{l_{S_2}}, \text{ with } C_{\gamma_{BN}} = \frac{R_B}{\rho_{S_2} l_{S_2}} \sqrt{\frac{3 T_{cS_1}}{\xi_{0S_2} T_{cS_2}}}. \quad (7)$$

Here, the quantities $C_\gamma$ and $C_{\gamma_{BN}}$ are independent of the thickness of the $S_1$ and $S_2$ films, because $\rho l$ is a material constant. The constant $C_\gamma$ depends only on the nature of the two materials involved, whereas $C_{\gamma_{BN}}$ also depends on the quality of the interface between the two films. For bilayers deposited under similar conditions and having only different film thickness, the same values of $C_\gamma$ and $C_{\gamma_{BN}}$ can be assumed. The dependence of the interface parameters on the film thickness can be determined by substituting the film thickness dependence of the mean free path into Eqs. (6) and (7). For complete electron scattering at the film surfaces the following equation for the mean free path l as a function of film thickness d holds[20]:

$$l = l_0 + l_0^2 / d \left\{ \frac{3}{2} [E_3(d/l_0) - E_5(d/l_0)] - \frac{3}{8} \right\}, \quad (8)$$

where the exponential integrals are defined by $E_n(x) = \int_1^\infty t^{-n} e^{-xt} dt$ and $l_0$ is the mean free path in the bulk material.

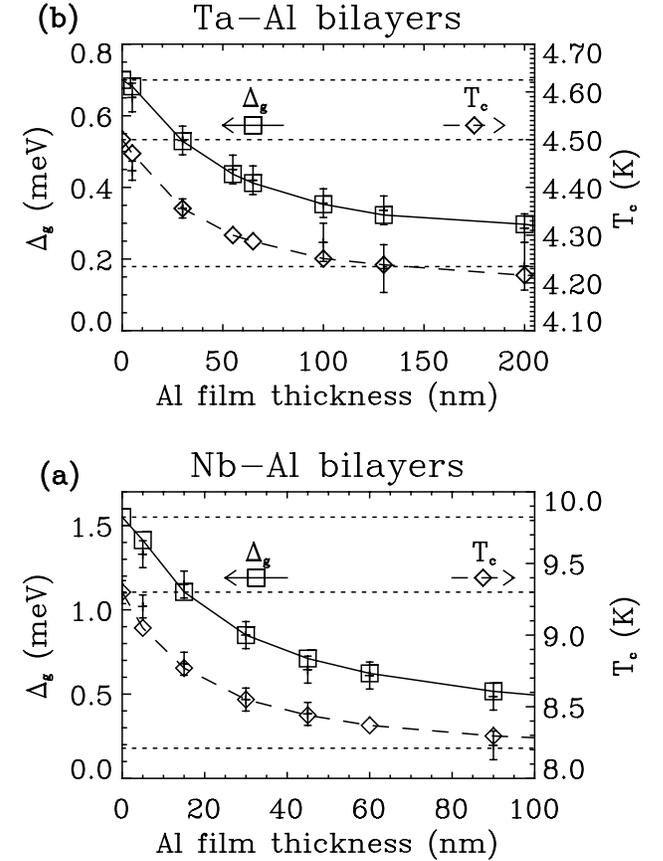

FIG. 2. Energy gap at 300 mK and $T_c$ as a function of Al film thickness for **(a)** Ta/Al and **(b)** Nb-Al bilayers. The Ta and Nb film thickness has a constant value of 100nm. Squares ($\Delta_g$, left scale) and diamonds ($T_c$, right scale) represent the calculated values from our model. The solid ($\Delta_g$, left scale) and dashed ($T_c$, right scale) lines are a guide to the eye between the calculated points. Crosses with error bars represent the corresponding experimental values. The dotted lines represent the bulk energy gap of Nb, Ta, Al and the bulk $T_c$ of Nb and Ta.

Analytical solutions for $T_c$ are possible only under certain limitations on layer thicknesses and interface parameters $\gamma$ and $\gamma_{BN}$, as discussed earlier in literature[7-18]. In order to calculate $T_c$ in the general case, we have solved the set of Eqs. (1) and (2) numerically. $T_c$ is defined as the maximum temperature for which nontrivial solutions for the pair potentials $\Delta_{S_1}$ and $\Delta_{S_2}$ exist.

The devices studied in this work are symmetrical $S_1S_2IS_2S_1$-junctions, where every electrode is made out of a superconducting bilayer $S_1S_2$. Two different kinds of devices are available, Ta-Al and Nb-Al based devices. The layers are deposited in a UHV-system. First a 100 nm thick layer of epitaxial Nb or Ta is laid down. Without breaking vacuum a polycrystalline Al film is then deposited on which a 10 Å Al-oxide barrier is grown. Then another polycrystalline Nb-Al or Ta-Al bilayer having the same film thickness as the base electrode is deposited on top of this oxide barrier. The thickness of the Al film depends on the sample and is varied between 5 and 200 nm. Details on material characteristics like bulk mean free path, residual resistance ratio and bulk coherence length of the different films can be found in Ref. 21. All values given in the following are averages between the values for the top and base electrode.

For every series of depositions (Ta or Nb based) the values of $C_\gamma$ and $C_{\gamma_{BN}}$ are determined experimentally. The determination is based on a comparison between experimental and simulated values of the energy gap at 300 mK and the $T_c$ of the bilayer. For the devices with 30 nm of Al from a deposition series a number of simulations of the energy gap at 300 mK and the $T_c$ with different values of $\gamma$ and $\gamma_{BN}$ are made. The model for the calculation of the energy gap of a superconducting bilayer is presented in Ref. 22. To a single pair of $\gamma$ and $\gamma_{BN}$ values corresponds a single pair of values of the gap at 300 mK and $T_c$[21]. By applying Eqs. (6) and (7), the interface constants $C_\gamma$ and $C_{\gamma_{BN}}$ can be determined. These were found to be equal to $C_\gamma = 0.4806$, $C_{\gamma_{BN}} = 0.6754$ nm$^{-\frac{1}{2}}$ for the Ta-Al bilayers and $C_\gamma = 1.372$, $C_{\gamma_{BN}} = 0.642$ nm$^{-\frac{1}{2}}$ for the Nb-Al bilayers. Knowing the values of $C_\gamma$ and $C_{\gamma_{BN}}$ and using Eq. (8) we can now determine the interface parameters for the whole Al thickness range, leaving the Ta and Nb thickness constant at 100 nm. The results are shown in Fig. 1.

With the interface parameters from Fig. 1 as input parameters to our model, we can calculate the gap at 300 mK and the $T_c$ of all different bilayers. The results of our simulations are shown in Fig. 2(a) for the Ta-Al bilayers and Fig. 2(b) for the Nb-Al bilayers. The agreement between theory and experiment is very good.

A model for the determination of the $T_c$ of superconducting bilayers was presented. The model is valid in the dirty and in the clean limit with the condition of diffusive scattering at the film boundaries, with no restrictions to film thicknesses, $T_c$ of the layers and resistivity of the superconductor-superconductor interface. Taking a lay-up of 100 nm of Ta or Nb topped with 30 nm of Al as a starting point in order to determine the interface parameters, we were able to predict the energy gap and the $T_c$ of a whole range of bilayers with different Al film thicknesses. The calculated values were compared to experimental values from Ta-Al and Nb-Al bilayers with Al film thicknesses ranging from 5 to 200 nm. The agreement between theory and experiment is very good for both the energy gap and the $T_c$.

[1] P. A. J. de Korte, *Nucl. Instr. and Meth. A* **444**, 163 (2000).
[2] P. Verhoeve, *Nucl. Instr. and Meth. A* **444**, 435 (2000).
[3] N. Rando et al., SPIE Proceedings **3435**, 74 (1998).
[4] C. A. Mears et al., *Nucl. Instr. and Meth.* A **370**, 53 (1996).
[5] K. D. Irwin, G. C. Hilton, D. A. Wollman, J. M. Martinis, *Appl. Phys. Lett.* **83**, 3987 (1998).
[6] K.A. Delin and A.W. Kleinsasser, *Supercond. Sci. Technol.* **9**, 227 (1996).
[7] P. de Gennes, *Rev. Mod. Phys.* **36**, 225 (1964).
[8] J. J. Hauser, H. G. Thenerer and N. R. Werthamer, *Phys. Rev.* **136**, 637 (1964).
[9] W. Silwert and L. N. Cooper, *Phys. Rev.* **141**, 336 (1966).
[10] W. L. McMillan, *Phys. Rev.* **175**, 537 (1968)
[11] B.Y. Jin and J. B. Ketterson, *Adv. Phys.* **38**, 189 (1989).
[12] K. D. Usadel, *Phys. Rev. Lett.* **25**, 507 (1970).
[13] A. A. Golubov, M. Yu. Kupriyanov, V. F. Lukichev, and A. A. Orlikovskii, *Sov. J. Microelectronics* **12**, 355 (1983).
[14] Z. Radovich, M. Ledvij and L. Dobrosavljevic-Grujic, *Phys. Rev. B* **43**, 8613 (1991).
[15] A. A. Golubov, Proc. SPIE'94, I. Bozovic, ed., Vol. **362**, p. 353 (1994).
[16] M. G. Khusainov, *JETP Lett.* **53**, 579 (1991).
[17] Ya. V. Fominov and M. V. Feigelman, *Phys. Rev. B* **63**, 094518 (2001).
[18] J. M. Martinis *et al.*, Nuclear Instr. *Methods in Phys. Res. A* **444**, 23 (2000).
[19] M. Yu. Kupriyanov and V. F. Lukichev, Sov. *Phys. JETP* **67**, 1163 (1988).
[20] D. Movshovitz and N. Wiser, *Phys. Rev. B* **41**, 10503 (1990).
[21] G. Brammertz, A. A. Golubov, A. Peacock, P. Verhoeve, D. J. Goldie, R. Venn, *Physica C* **350**, 227 (2001).
[22] G. Brammertz, A. Poelaert, A. A. Golubov, P. Verhoeve, A. Peacock, H. Rogalla, *J. Appl. Phys.* **90** 1, 355 (2001).